\documentclass[Afour,sageh,times]{sagej}

\usepackage{moreverb,url}

\usepackage[colorlinks,bookmarksopen,bookmarksnumbered,citecolor=red,urlcolor=red]{hyperref}

\newcommand\BibTeX{{\rmfamily B\kern-.05em \textsc{i\kern-.025em b}\kern-.08em
T\kern-.1667em\lower.7ex\hbox{E}\kern-.125emX}}

\begin{document}

\runninghead{Smith and Wittkopf}

\title{Ethical Leadership in the Age of AI: Challenges, Opportunities and Framework for Ethical Leadership}

\author{Udaya Chandrika Kandasamy, WGU School of Business, USA}

\begin{abstract}
Artificial Intelligence (AI) is currently and rapidly changing the way organizations and businesses operate. Ethical leadership has become significantly important since organizations and businesses across various sectors are evolving with AI. Organizations and businesses may be facing several challenges and potential opportunities when using AI. Ethical leadership plays a central role in guiding organizations in facing those challenges and maximizing on those opportunities. This article explores the essence of ethical leadership in the age of AI, starting with a simplified introduction of ethical leadership and AI, then dives into an understanding of ethical leadership, its characteristics and importance, the ethical challenges AI causes including bias in AI algorithms. The opportunities for ethical leadership in the age of AI answers the question: “What actionable strategies can leaders employ to address the challenges and leverage opportunities?” and describes the benefits for organizations through these opportunities. A proposed framework for ethical leadership is presented in this article, incorporating the core components: fairness, transparency, sustainability etc. Through the importance of interdisciplinary collaboration, case studies of ethical leadership in AI, and recommendations, this article emphasizes that ethical leadership in the age of AI is morally essential and strategically advantageous.
\end{abstract}

\maketitle

\section{Introduction}
Organizations and businesses across the globe are increasingly integrating Artificial intelligence (AI) into their operations. AI technologies, ranging from Machine Learning and Natural Language Processing to Robotics and Automation are reshaping businesses and organizational operations at an exponential pace. AI has tremendous benefits like increasing efficiency, improving decision-making and driving innovation. (Kaplan, 2016). However, they also present significant challenges like algorithmic bias, privacy concerns, and possible job displacements.  The implications of AI can be long-lasting, affecting not just the organizations and businesses integrating these technologies, but can also impact societal norms and values. 
Ethical leadership plays a crucial role and takes over a new dimension as AI rapidly transitions the fundamental way businesses operate. (Mendonca and Kanungo, 2007). Within the AI context, ethical leadership expands beyond regulatory compliance and involves exploring and solving complex moral dilemmas that arise from the use of AI algorithms and automated decision-making systems. Ethical leadership emphasizes ethical considerations as AI is deployed and managed by organizations. Ethical considerations are essential to make sure that AI systems operate fairly and responsibly. 
Leaders of organizations have the potential to influence and implement environments that promote ethical practices, encourage openness about AI implications, and engage employees and stakeholders in communications about values and ethics. Through ethical leadership in the age of AI, organizations can analyze, manage and solve challenges, uncover opportunities and maximize benefits presented by AI technologies and contribute to fair, equitable and just societies. This article explores ethical leadership in the age of AI, challenges faced while integrating and using AI ethically, opportunities for leaders in influencing ethical AI practices, and proposes a framework for ethical leadership in the age of AI. Additionally this article includes the importance of interdisciplinary collaboration, case studies of ethical leadership practices in AI and recommendations for leaders.

\section{Understanding Ethical Leadership}
Ethical leadership is the practice of leading with a strong moral compass, prioritizing the well-being of stakeholders and employees, and fostering a culture of fairness, integrity, transparency, and accountability. Ethical leadership is not only about guiding organizations through technological change but also ensuring that organizational changes align with societal values and norms. Ethical leaders know what is the ethical thing to do and they make this the foundation for their leadership through their actions. (Brown and Treviño,2006;Buck, 2023). Employees and other organizational members tend to follow leaders. When ethical leaders behave morally and follow ethical standards, they adapt to and exhibit those traits themselves. (Brown and Treviño,2006).
Ethical leaders practice unique sets of characteristics that distinguish them from their peers. These characteristics enable them to navigate the complexities of AI implementation as they maintain their commitment to ethical standards. Top characteristics of ethical leaders include:
\begin{itemize}
\item {Integrity:  Ethical leaders exhibit integrity by being honest and truthful. Ethical leaders consistently align their actions with their values and adhere to ethical principles. They set an example for others by making decisions that reflect integrity, even when faced with challenges or pressures to do otherwise. Ethical leaders empower employees and create a culture of ownership and accountability through integrity. (Buck, 2023).
}
\item {Empathy: Ethical leaders understand and consider the perspectives and feelings of their stakeholders and employees. They practice empathy through active listening and communication. Empathy allows them to engage effectively with all members of the organization, address concerns, and build a supportive environment where diverse viewpoints are valued. (Buck, 2023).
}
\item {Accountability: Ethical leaders take responsibility for their decisions. In case the decisions lead to unintended outcomes, ethical leaders hold themselves accountable. By fostering a culture of accountability within their organizations, they encourage employees to be accountable, and create opportunities for them to own their actions and learn from mistakes. (Buck, 2023).
}
\item {Transparency: Transparency in ethical leadership is about open communication. Ethical leaders who are transparent about their intentions and decision-making processes build trust and create an environment where employees feel safe to voice their opinions and concerns. By being transparent, ethical leaders contribute to the alignment of organizational operations with its vision, mission, and goals. (Hartog, 2015).
}
\item {Fairness: Ethical leaders exhibit fairness by valuing and treating everyone equally. Fairness eliminates biases. Ethical leaders promote a culture of inclusion through fairness and equality. (Hartog, 2015).They remain uninfluenced by biases and make decisions based on objective criteria.
}
\item {Vision: A major characteristic of ethical leaders is the ability to visualize the future of organizations clearly, that incorporates ethical considerations and impact on the society. (Hartog, 2015).They serve as an inspiration to many through this ethical leadership characteristic that prioritizes business success as well as contributes to the greater good. 
}
\item {Trustworthy: Ethical leaders are trustworthy and excel in building trust through consistent actions, transparency, and open communication. Employees engage openly and contribute to the organization’s success when they trust their leaders. Ethical leaders also build trust amongst stakeholders, customers and partners, thereby enhancing the organization's reputation. A good foundation of trust allows leaders to seek input and work with their teams, promoting a collaborative approach to ethical decision-making. (Van Den Akker et al.,2009)
}
\item {Resilience: Ethical leaders practice resilience by adapting to uncertainty and working through solutions. When faced with challenges, ethical leaders showcase the ability to bounce back from those challenges, learn and respond efficiently. (Crews, 2015).
}
\item {Respect: Ethical leaders demonstrate respect by valuing everyone and taking factors similar to diversity, opposing views and contributions into consideration. (Buck, 2023).
}
\end{itemize}
Why is ethical leadership important? Ethical leadership impacts organizations and eventually societies and the results can be phenomenal. (Ciulla JB and Ciulla JB, 2020). Improved employee morale and engagement, a strong and positive organizational culture, increased team efficiency are noteworthy impacts from ethical leadership. Another impact of ethical leadership is organizational reputation. How a business is perceived b y its customers is influenced by ethical leadership. An organization that remains committed to ethical values will retain loyal customers and attract new customers. Ethical leadership is important in attracting and retaining the right talent. Ethical leaders establish a positive workplace culture and employees seek workplaces that value them and show a vested interest in their well-being and growth. Ethical leadership boosts investor relations and increases trust amongst partners and vendors. Without ethical leadership, organizations may seek short-term gains and will be driven by unethical decisions. Unethical leadership practices will lead to a weak organizational culture, loss of trust and loyalty amongst employees, customers and stakeholders. Without ethical leadership, organizations cannot face challenges and solve problems. Decision-making will be impacted and biases may influence the processes. Severe cases of absent ethical standards can lead to violation of law and legal complications resulting in lawsuits, penalties, loss of finances and damage to organizational reputation.

\section{Ethical Challenges Presented by AI:}
The integration of AI technologies into organizations and businesses brings several ethical challenges. Ethical leadership plays a vital role while organizations traverse through these challenges. Addressing the ethical challenges presented by AI ensures that ethical standards are met and that the integrated AI technologies align with societal norms and values and contribute to the greater good. Some ethical challenges associated with AI are: bias, privacy concerns, transparency and accountability, job displacement, environmental issues.
\begin{itemize}
\item {Bias: One of the most pressing ethical challenges presented by AI technologies is the potential for AI bias. Also known as algorithmic bias or machine learning bias. There are multiple types of AI biases. AI systems learn from historical data and large datasets known as big data. Historic data may contain biases. Usage of existing biases in historical data is historical bias. Existing biases in data can perpetuate or even magnify unethical practices. (Holdsworth, 2023). A few other forms of AI bias include sample bias (bias from sample datasets that do not accurately represent real word data), label bias (incorrect labeling), confirmation bias (trusting data that confirms existing beliefs), evaluation bias (incorrect or insufficient data evaluation). (Seldon, 2021).
The implications of AI bias are many. Biased algorithms can result in poor hiring processes and unfair hiring practices, leading to favorable outcomes for candidates from certain demographics over others. Missing data or underrepresented data can result in skewed results. Biased AI algorithms may provide incorrect supply demand forecasts that can hurt businesses.   Such biases can lead to loss of reputation, loss of trust in the organization, lost opportunities for specific groups of people and may even lead to regulatory issues. (Eitel-Porter R, 2021)
To address bias presented by AI and to mitigate the risks caused by them, ethical leaders must encourage the use of diverse and representative datasets, continuous and rigorous checking for bias in algorithms, and implement fairness assessments. Implementation of and continuous monitoring and auditing of AI systems can identify and reduce biases, and in the long run may solve biases, ensuring AI algorithms are aligning with ethical principles.
}
\item {Transparency and Accountability: AI algorithms are constantly improvising and becoming sophisticated, and as they do so, their decision making processes are getting complex and less transparent, making it difficult to understand how AI systems are reshaping organizations. The “Black Box” nature of AI systems refers to the lack of transparency of how some AI models make decisions. (Eitel-Porter R, 2021). This poses significant challenges regarding transparency and accountability. AI algorithms based on complex machine learning models, often function in ways that are not easily understandable to end users or stakeholders. The “Black Box Problem” raises ethical concerns about how decisions are made and who is held accountable for those decisions. For instance, if an AI system denies a loan application without sufficient explanation and objective evidence, understanding how AI made the decision and the rationale behind the decision can be challenging for both the applicant and the financial institution. 
	There are multiple implications caused by this ethical challenge. Black box AI models are vulnerable to attacks and can weaken security of software systems. Attackers can exploit the vulnerabilities found in AI systems and manipulate the results of AI systems. With reduced transparency, there is reduced trust in the AI systems. It is hard to troubleshoot black box models. It is difficult to hold someone responsible for the decisions made by AI, caused by the lack of accountability. Another implication is the lack of clarity on accuracy validation. No transparency makes it close to impossible to test and validate results from a black box. Use of such models is ethically concerning in sectors like finance and healthcare, where organizations being transparent and accountable is crucial.
To address the “Black Box Problem” and to promote transparency, ethical leadership must emphasize on the commitment to transparency and accountability while using AI. Organization leaders should mandate clear documentation of algorithms, decision-making processes, continuous monitoring and fixing of flaws and analyze all factors that influence AI outcomes. Establishing organization-wide processes and mechanisms for accountability, where individuals or teams are responsible for decisions made by AI is another ethical decision leaders can implement to mitigate the risks caused by the lack of transparency and accountability of AI. (Walmsley, 2021).
}
\item {Privacy Concerns: Extensive collection of data and analysis of datasets is a fundamental block of AI systems. Businesses and organizations using AI integration gather data for analysis and to train their AI algorithms. These data collection and analysis practices raise privacy concerns. Privacy concerns occur in several forms. Intentional or unintentional misuse of data, unauthorized access of sensitive information, unclear data ownership and violation of the basic right to privacy are key privacy concerns. 
The implications of data collection practices and how data is used must be carefully considered to protect individual rights. Misuse of data can take several forms and may lead to identity theft, discriminatory practices, unauthorized monitoring or surveillance of individuals. There is a high risk of cyber attacks leading to unauthorized access of sensitive information. For instance, breaches to a healthcare organization using AI can expose personal information of individuals as well as the organization, leading to loss of trust in the system and opportunities. Though several organizations include informed consent in their data collection practices, who owns the data collected is often ambiguous. Individuals may feel a sense of ownership over their data, however organizations retain the rights to collected data. The right to privacy is constantly challenged with the rise of AI. Ethical dilemmas arise from not completely understanding how data will be used through informed consent data collection practices. (Yaou and Hyounae, 2023).
To reduce the privacy concerns and the risks surrounding them, ethical leaders must encourage organizations to implement ethical data collection and usage practices. Compliance with regulations such as GDPR (General Data Protection Regulation) and prioritizing robust data governance frameworks are essential. These practices include providing transparency about what data is collected, the purpose behind its collection,and how it will be used. (Judge et al.,2024;Bastit B et al., 2023). Ethical leaders can empower the users and customers to manage their data and preferences by including opt-in and opt-out options and ways to delete their data. Implementing ethical auditing practices and stakeholder engagement are other ethical leadership practices to mitigate privacy concerns.
}
\item {Job Displacement:  Automation of tasks through AI is revolutionizing industries and reshaping the workforce across multiple sectors. While AI looks promising in increasing efficiency, productivity, and cost savings, it also raises several ethical considerations regarding job displacement and the resulting changes in employment dynamics. (Frey and Osborne, 2017). Organizations and leaders must navigate these complexities to ensure a just and equitable transition and opportunities for workers. 
The implications from job displacement may be severe. The possibility of widespread job displacement presents numerous ethical dilemmas for organizations. Employers face a moral obligation to consider the consequences to their workforce due to AI integration into their organizations. Several workers may find themselves unprepared for such shifts in jobs, creating disparities in employment opportunities and income. Job displacements can eventually affect the families of workers and communities as a whole. Inequalities may deepen with uncertainty in job markets. Individuals without access to the right resources may not transition into new roles, leading to unemployment and underemployment. Economic imbalances and divides between those who can adapt to new technologies and those who cannot can occur, leading to entrenching social and economic inequalities. (Tennin et al.,2024).
To ethically manage the impact of automation and to handle job displacement, organizations and its leaders must take ethical steps. Ethical leadership in the age of AI requires proactive measures to address negative impact. Investing in reskilling and upskilling programs to help workers transition to new roles that complement AI technologies and establishing a culture of lifelong learning are actions ethical leaders can implement. Adaptability within organizations can empower employees to thrive in the age of AI. Partnering with educational institutions for learning, transparent communication, providing support systems and resources are other ethical measures to address this ethical challenge. 
}
\item {Environmental Issues:  Environmental issues and issues with environmental sustainability is another ethical challenge from the use of AI. Critical issues include resource consumption, waste management, and the broader impact of AI technologies on our planet. Some AI systems require extensive computational power and demand significant energy resources. Training models consume large amounts of electricity and most of this is non-renewable energy. Carbon emissions are increased with such consumptions and climate change is accelerated as a result.  (Brevini, 2020).
	Improper waste management is another environmental concern. The rapid advancement of AI technologies leads to increased electronic waste. Obsolete and discarded electronics contain harmful substances that leach into the environment and pose ecological risks. Other than energy, physical space consumption is another implication. Large amounts of land and water are consumed in the process of expansion of data centers for the usage of AI systems. AI systems when used in sectors like agriculture and mining can lead to over exploitation of natural resources. They may increase profits and yields in the short run, however natural resources may be depleted if sustainability is not practiced. If AI algorithms are designed to prioritize profit over sustainability, organizations may overlook the ecological consequences leading to decisions that harm the environment. (UN environment programme, 2024)
	Ethical leadership can address the environmental challenges in numerous ways. Ethical leaders should prioritize the development and implementation of AI technologies that emphasize sustainability. Investing in green technologies, optimizing algorithms for energy efficiency, and using renewable energy sources in data centers and AI operations are some means to accomplish sustainability. Ethical leaders can lead by example by establishing environmental standards within the organization and by promoting awareness, transparency and accountability regarding the environmental impact of AI systems. Investing in research and development, engaging stakeholders and customers on sustainability and waste management, being responsible about environmental impact, establishing sustainable policy changes within the organization are other recommended ethical practices for leaders in addressing the environmental concerns ethical challenge. (Brevini, 2020; UN environment programme, 2024).
}
\end{itemize}

\section{Opportunities for Ethical Leadership in the age of AI:}
Ethical leadership in the age of AI is a guiding principle for organizations traversing the challenges of AI and is also a catalyst for unlocking positive and exciting opportunities. Organizational leaders have the opportunity to leverage AI to enhance the organizational performance and operations and also to sustain ethical ethical standards and societal values and norms. Five key opportunities for ethical leadership in the AI era are discussed in this section: fostering innovation, enhancing decision-making, building trust, engaging stakeholders and employees, and influencing policies and regulations.
\begin{itemize}
\item {Fostering Innovation: Ethical leadership can play a critical role in driving responsible use of AI for fostering innovation. By advocating and championing ethical practices in AI development, organizational leaders can create an environment where innovation flourishes while addressing ethical concerns. The integration of ethical considerations in the design and development of AI technologies is another mechanism to foster innovation. Promoting diverse teams that bring varied perspectives and experiences to the innovation process can help identify potential biases and ethical implications early on. Investing in research and development that focuses on ethical AI applications can lead to innovative solutions that prioritize social good. For instance, developing AI systems that enhance accessibility for people with disabilities exemplifies how ethical leadership can steer innovation toward addressing societal challenges. Fostering a culture that encourages experimentation and calculated risk-taking while ensuring that ethical implications are always considered is another ethical leadership opportunity to foster innovation. With the support of ethical leaders, organizations can generate innovative solutions that align with ethical standards, by promoting an environment where employees feel safe to explore new ideas. (Mendonca and Kanungo, 2007).
}
\item {Enhancing Decision-Making: AI technologies can significantly improve decision-making processes and aid in creating new and better decision-making processes within organizations. Ethical leaders can harness this potential to drive better outcomes by leveraging AI generated data-driven insights, scenario analyses, and by encouraging collaborative decision-making.  AI systems have the ability to analyze large datasets and provide insights that help with strategic decisions. (Fountaine et al.,2019). Ethical leaders can utilize these insights to make informed choices regarding business objectives by considering ethical implications. AI allows simulation of scenarios and prediction of potential outcomes, allowing for more refined decision-making. These capabilities can be used by ethical leaders to assess the impact of their strategic decisions on employees, stakeholders, and customers, ensuring that those choices align with ethical principles. Another unique way ethical leadership can maximize this opportunity is by encouraging collaborative decision-making. Diverse opinions and views can be involved in critical discussions by leveraging AI. This opportunity for ethical leaders enhances the quality of decisions and fosters a sense of ownership, transparency and accountability among team members.}
\item {Building Trust: A key characteristic of ethical leaders is being trustworthy. And a key opportunity in the age of AI for ethical leaders is building trust. By combining their key characteristics and by creating a transparent culture around AI use, ethical leaders can build trust within organizations and with external stakeholders. Ethical leaders can take proactive steps in this direction through clear communication, by establishing ethical guidelines and by engaging in open dialogue. A commitment to clear communication regarding how AI technologies are used within the organization will help the organization and related people in getting comfortable with AI. (Desimone, 2024). Providing stakeholders with insights regarding AI decision-making processes and data usage builds confidence in the integrity of the systems in place. Developing and distributing ethical guidelines for using AI ethically within the organization helps establish a framework that promotes responsible practices. Active involvement of employees in the creation of these guidelines reinforces a collective commitment to ethical standards. Establishing a safe space for open conversations about the ethical implications of AI fosters a culture of trust. Additionally, ethical leaders can create forums where employees and stakeholders can express concerns, ask questions, and engage in discussions about the ethical dimensions of AI technologies.}
\item {Engaging Stakeholders and Employees: Engaging employees and stakeholders in AI governance is a primary responsibility of ethical leadership. By engaging stakeholders in the decision-making process, leaders can ensure that diverse perspectives are considered. The stakeholders opinion and perspectives of the organization's usage of AI will be an addon from engaging stakeholders. (Tennin et al.,2024). Ethical leaders can implement collaborative governance models to include representatives from various stakeholder groups and expand it to employees, customers, communities and eventually societies. This promotes inclusivity, diversity and allows for a broader understanding of the ethical implications and opportunities of AI technologies. Establishing robust feedback mechanisms enables easy access to voice their concerns and experiences with AI systems. Ethical leaders can use this feedback to make adjustments and improvements, ensuring that the organization and the AI technologies used within serve the interests of all stakeholders. To further boost engagement, providing education and resources about AI technologies, empowers stakeholders to engage meaningfully in discussions about governance. Facilitating workshops, training sessions and establishing a culture of continuous learning that equip employees and customers with the knowledge to contribute to AI governance are other opportunities for ethical leaders.
}
\item {Influencing Policies and Regulations: Ethical leadership has a major opportunity in the age of AI by influencing policies and regulations. Ethical leaders can impact how AI is integrated into society, ensuring that its deployment aligns with ethical considerations and public welfare. By advocating robust and fair privacy regulations that limit the collection, storage and use of data by AI systems and by championing privacy, ethical leaders can establish safe AI usage. Leaders can protect organizations and individuals by creating strict security standards to govern development and deployment of AI, developing protocols to safeguard AI enabled systems by collaborating with cybersecurity experts. AI is a global phenomenon. Leaders can collaborate with other leaders globally and engage with international organizations to shape global ethical standards. (Judge et al.,2024). This will result in better regulations and consistent application of AI across sectors. Educating people about the benefits, risks, policies and regulations surrounding AI systems is another opportunity for ethical leaders to influence safer policies and regulations. 
}
\end{itemize}

\section{Framework for Ethical Leadership:}
A framework for ethical leadership in the age of AI is an essential and effective way for organizations and its leaders to traverse the complexities and challenges brought by AI and maximize the opportunities that AI technologies bring with them. The design of this framework is built on five key fundamental principles that guide the responsible and ethical use of AI technologies and is distinguished by the acronym {\textit{(AFPTS): Accountability, Fairness, Privacy, Transparency, Sustainability}. Other than a key set of guiding principles, the AFPTS framework also outlines actionable strategies for ethical leaders to implement and promote ethical practices. Furthermore, fostering interdisciplinary collaboration is essential for addressing the multidimensional challenges posed by AI. Let us review the key principles of AFPTS framework for ethical leadership in the age of AI with actionable strategies and benefits: 

\subsection{AFPTS framework - Key Principles, Actionable Strategies and Benefits:}
\begin{itemize}
\item {Accountability: Accountability in ethical leadership in the age of AI ensures that organizations are held responsible for their AI systems’ actions and impacts. Clear accountability structures regarding AI’s decisions and actions must be established by organizational leaders. Identifying individuals or teams responsible for the ethical implications of AI systems, establishing mechanisms in place to address and rectify any issues or harm caused by AI deployment, allowing individuals to seek recourse, and promoting ethical standards, and compliance with regulations are vital actionable strategies for organizations to operate ethically and be accountable. Accountability improves trust and keeps up the organizational reputation among users and stakeholders. (Dignum, 2017). Companies known for responsible AI practices attract more customers and partners, enhancing their market position and potentially increasing revenue and maximizing profit. Accountability establishes a culture of continuous improvement. Continuous improvement encourages ongoing evaluation and enhancement of AI systems, so they stay aligned with ethical standards and societal values. By implementing accountability measures and continuous improvement, organizations can minimize the risk of lawsuits and regulatory penalties, leading to significant cost savings.}
\item {Fairness: Fairness is a key principle to ensure AI systems operate without bias and is a primary building block to promote equity. (Hartog, 2015). Leaders must prioritize fairness in algorithm design and data selection, development, and actively work to identify and mitigate potential biases that could lead to discriminatory outcomes. Bias mitigation is another actionable strategy for ethical leaders. Bias mitigation reduces biases in AI algorithms, ensures equitable treatment across different demographics. Preventing discrimination whether intended or unintended and promoting social justice are major benefits for organizations practicing fairness. Inclusivity is another benefit when fairness is fostered at workplaces. Inclusivity encourages the development of AI that considers diverse perspectives and needs, leading to more comprehensive and effective solutions. Talent attraction, management and retention for organizations practicing fairness will be efficient through these actionable strategies. Users and communities tend to trust organizations that strive to build trust and keep it up. Ethical leadership can significantly contribute to trust building, customer loyalty, customer retention rates and increased lifetime value by advocating fairness when using AI. A primary benefit from this key principle is market expansion. AI systems when biases are mitigated and when inclusivity and diversity is implemented, can reach underserved markets by ensuring equitable access and treatment, opening new revenue streams.}
\item {Privacy: Protecting the privacy of individuals is of great importance in the era of AI. (Yaou and Hyounae, 2023). Ethical leaders must establish mechanisms to ensure that their organizational data collection practices respect users' rights and comply with relevant regulations. Actionable strategies include implementing robust data protection measures and obtaining informed consent for data usage. User data protection strategies assure end users that their personal data is safeguarded by the organization. Ethical leaders must prioritize privacy in every section of their organizations and allow development and implementation of data privacy strategies to respect individuals’ rights and enhance their control over their information. Users adopt and accept AI technologies that prioritize their privacy. Ethical leaders, through privacy and these actionable strategies can further build confidence and trustworthiness in the AI systems used by their organizations. Another vital aspect of privacy in the age of AI is regulatory compliance. Ethical leadership should mandate adherence to laws and regulations related to data protection in order to mitigate legal risks for organizations. Organizations respecting user privacy can lead to higher adoption rates of AI products, as consumers are more likely to engage with companies that protect their personal data. Organizations' cost efficiency improves through privacy. Privacy implementation minimizes data breaches and the associated costs similar to fines and remediation. 
}
\item {Transparency: Transparency is critical for encouraging trust in AI technologies. Ethical leaders must advocate for clear communication about how AI systems function, the data used for training, and the decision-making processes involved. Openness allows understanding and scrutinizing AI operations, and enhances confidence in the technology.  (Dignum, 2017). Actionable strategies include publishing organizational news and articles focused on their usage of AI, requesting informed consent for fair data collection and usage. This helps the audience gather better understanding of AI systems and to comprehend how decisions are made, which is necessary for trust and acceptance. Informed consent empowers users by providing clear information about data usage and AI functionalities. Another actionable strategy is to establish streams to gather feedback from users and stakeholders regarding their AI usage. Feedback allows transparency by letting participants be anonymous, additionally a collaborative approach in refining AI systems is established. Transparency leads to improved insights and decision-making, enhancing efficiency of operations. When insights are better and operational efficiency increases, costs are reduced. Ethical leaders by implementing transparency, garner competitive advantage to their organizations. Companies that are transparent about their AI practices differentiate themselves in the marketplace. Attracting more customers and commanding premium pricing becomes possible through this key principle. }
\item {Sustainability: Ethical AI use should consider the long-term impact on society and the environment. Leaders have the authority to promote sustainable practices in AI development, ensuring that technologies designed are focussed on minimizing ecological footprints and promoting social well-being. (Kang, 2019). Actionable strategies include environmental consideration, assessing long-term viability and being socially responsible in every organizational sector using AI. Environmental consideration encourages the development of AI solutions that reduce environmental impact and promote a healthier planet. Assessing long-term viability focuses on creating systems that are effective and sustainable in the long run. Ensuring that future generations benefit from ethical AI advancements must be a key component in such assessments. Sustainable AI practices lead to more efficient resource use,better recycling processes and efficient waste management. Operational costs will be reduced for the organization through sustainability. Several investors prioritize ESG (Environmental, Social, and Governance) criteria. Ethical leadership, through strong sustainability practices may easily attract investment and funding.
}
\end{itemize}
To effectively implement these principles, ethical leadership can adopt a range of strategies that promote ethical AI practices within their organizations:
\begin{itemize}
\item {Developing Ethical Guidelines: Leaders must establish comprehensive ethical guidelines for AI use that outline the organization’s commitment to ethical practices. Those guidelines should be communicated clearly and integrated into all aspects of AI development,deployment,usage and maintenance.}
\item {Training and Education: Ongoing training and education on ethical AI principles is essential for fostering a culture of responsibility. Leaders can implement training programs and learning  management systems that educate employees about the ethical implications of AI and empower them to make informed decisions. (Fountaine et al.,2019)}
\item {Establishing Ethics Review Boards: Ethics review boards or committees can help organizations assess the ethical implications of AI projects. These boards can evaluate proposed AI initiatives, ensuring alignment with ethical standards and principles before implementation.}
\item {Promoting Open Dialogue: Encouraging open discussions about ethical dilemmas related to AI allows and eventually establishes a culture of transparency and accountability. Leaders can create forums where employees and stakeholders can voice concerns, share insights, and engage in collaborative problem-solving.}
\item {Implementing Continuous Monitoring: Ethical AI practices require ongoing assessment and improvement. Leaders should establish mechanisms for monitoring AI systems after deployment to ensure compliance with ethical standards and to identify any unintended consequences.}
\end{itemize}

\section{The Importance of Interdisciplinary Collaboration:}
The complexities associated with AI necessitate interdisciplinary collaboration among various stakeholders. Ethical leaders play a central role in initiating and establishing interdisciplinary collaboration. Collaboration between technologists and ethicists, engagement with business leaders within, involving diverse stakeholders and investors, partnerships with academic and research institutions are means for interdisciplinary collaboration. (Fountaine et al.,2019)
\begin{itemize}
\item {Collaboration Between Technologists and Ethicists: Technologists bring technical expertise to AI development and usage. Ethicists provide insights into the moral implications of these technologies. Establishing teamwork between these groups and by allowing them to work together, ethical leadership can identify potential ethical issues early on in the design process and develop solutions that prioritize both functionality and ethics.}
\item {Engagement with Business Leaders: Business leaders play a crucial role in aligning ethical AI practices with organizational goals. Collaboration with business leaders ensures ethical considerations are integrated into strategic planning and decision-making processes.}
\item {Involvement of Diverse Stakeholders and Investors: Engaging a diverse range of stakeholders and investors, including employees, customers, and community representatives diversifies the ethical framework of AI initiatives. By being open and incorporating multiple perspectives, organizations can identify potential biases and blind spots, fostering a more inclusive approach to AI development.}
\item {Partnerships with Academic and Research Institutions: Collaborating with academic and research institutions can provide organizations with access to cutting-edge research and best practices in ethical use of AI. These partnerships can help organizations stay informed about emerging trends and innovations while reinforcing their commitment to ethical leadership.}
\end{itemize}
A well-defined framework for ethical AI leadership is essential for guiding organizations in the responsible use of AI technologies. By adhering to principles of accountability, fairness, privacy, transparency, and sustainability, leaders can create a culture of ethical AI practices. Furthermore, fostering interdisciplinary collaboration will enhance the effectiveness of these initiatives, enabling organizations to use AI with integrity and social responsibility. This framework will empower leaders to harness the transformative potential of AI, while ensuring that ethical considerations play a central role in their strategies.

\section{Case Studies of Ethical Leadership in AI:}
Some organizations have demonstrated ethical leadership in the age of AI and across several sectors.  Notable case studies of such organizations showcase innovative approaches to integrating ethical considerations into AI development and deployment.
\begin{itemize}
\item {Microsoft has established itself as a pioneer in ethical AI through its AI Ethics Committee and AI for Good initiative. They emphasize fairness, accountability, transparency, and inclusiveness in their AI systems. Their comprehensive AI ethics framework guides product development and external partnerships.}
\item {Google’s AI Principles articulate their commitment to developing AI responsibly. The organization has implemented rigorous internal review processes for AI projects, focusing on the potential societal impact. Google collaborates with external stakeholders to improve its understanding and application of ethical AI.}
\item {IBM has long advocated for ethical AI through their AI Fairness 360 Toolkit, which helps organizations detect and mitigate bias in AI systems. Their commitment to transparency and accountability is evident in their public documentation of AI model decision-making processes.}
\item {Salesforce has integrated ethical considerations into its AI offerings through their Einstein AI platform. The company has launched initiatives to promote ethical use of AI in customer relationship management (CRM) while prioritizing data privacy and user consent.}
\item { OpenAI has set a benchmark for ethical considerations in AI research and deployment. With focus on safety and long-term impact, OpenAI engages in public dialogues about the implications of AI, aiming to align its objectives with broader societal values.}
\end{itemize}
These organizations have employed a variety of strategies to embed ethical considerations into their AI practices. Many have created comprehensive frameworks that define ethical principles and provide guidelines for AI development. This structured approach ensures consistency across projects and aligns teams with ethical standards. Organizations like Google and Microsoft actively engage stakeholders to gather diverse perspectives on AI impact, fostering a collaborative environment for ethical discussions. IBM’s AI Fairness 360 Toolkit is a prime example of developing practical bias mitigation tools that empower organizations to address bias proactively. These tools facilitate the identification and reduction of bias, leading to fairer AI systems. Companies such as OpenAI and Salesforce prioritize transparency by sharing insights into their algorithms and decision-making processes. This openness helps build trust with users and stakeholders. The outcomes of these strategies have been largely positive, resulting in increased public trust, enhanced user satisfaction, and a competitive edge in the marketplace. By prioritizing ethical leadership, these organizations have positioned themselves as industry leaders and role models for responsible AI practices.
By learning from these exemplary organizations, other companies can navigate the ethical use of AI and establish a culture of responsible innovation that benefits both businesses and society at large.

\section{Recommendations for Leaders:}
As AI technology continues to evolve rapidly, leaders have a pivotal role in ensuring ethical considerations are seamlessly integrated into their organizations' AI strategies. Here are practical steps leaders can take to cultivate a culture of ethics within AI development and deployment.
\begin{itemize}
\item {Develop and establish an ethical AI framework that outlines the organization’s ethical principles regarding AI, including guidelines for fairness, accountability, transparency, and user privacy. Regularly revisiting and updating this framework can ensure it remains relevant as technology and societal expectations evolve.}
\item {Conducting ethical impact assessments regularly for AI projects can help identify potential risks and ethical dilemmas early in the development process. Leaders should create standardized protocols for these assessments to evaluate the implications of AI technologies on stakeholders, particularly marginalized communities. (Mantelero, 2022).}
\item {Fostering diverse teams that bring varied perspectives is crucial in developing ethical AI. Leaders should prioritize diversity in hiring practices and create cross-functional teams that include data scientists, ethicists, sociologists, and community representatives to encourage comprehensive discussions about ethical implications.}
\item {Establishing governance structures that oversee AI initiatives, ensuring compliance with ethical standards. This may involve forming ethics committees or appointing Chief Ethics Officers who can guide AI development and address ethical concerns proactively.}
\item {Engaging with external stakeholders including customers, advocacy groups, and regulatory bodies can provide valuable insights into ethical concerns. Leaders should facilitate open dialogues and workshops to better understand societal expectations and ethical challenges associated with AI.}
\item {Promoting continuous learning and adaptation in leadership by staying informed on AI developments, encouraging feedback and reflection and adapting to regulatory changes is a mindset leaders must embrace to keep up in the age of AI. (Fountaine et al.,2019)}
\item {Ethical leaders must strive to build a culture of ethics in AI development and deployment by modeling ethical behavior, implementing training programs and creating incentives for ethical practices, encouraging open dialogue and highlighting ethical success stories.}
\end{itemize}
By implementing these recommendations, leaders can foster a responsible culture and approach that aligns with societal values while driving innovation and success in their organizations.

\section{Conclusion:}
The future of ethical leadership in the age of AI is complex yet achievable and dynamic. The rapid advancement of AI technologies necessitates a proactive approach to ethical leadership. Leaders will face new challenges and opportunities as they lead organizations integrating AI. As AI systems increasingly influence multiple processes across various sectors, leaders must prioritize ethical considerations to ensure these technologies align with societal values and human rights. By embracing the key principles of AFPTS ethical AI framework, fostering interdisciplinary collaboration, and advocating for responsible governance, leaders can ensure that their organizations contribute positively to the ethical development and deployment of AI technologies. As we move forward with AI, the role of policy and regulation will be critical in shaping a future where ethical leadership is not just an option but a foundational principle guiding AI innovation.
In summary, while AI presents remarkable opportunities for innovation, efficiency and a sustainable future, it also poses significant ethical challenges that leaders must confront. By addressing issues of bias and fairness, promoting transparency and accountability, ensuring ethical data practices, and mitigating the effects of job displacement, ethical leaders can create a positive impact in the present and future. By fostering innovation, enhancing decision-making, building trust, and engaging stakeholders, leaders can guide their organizations toward responsible AI practices that align with ethical standards. Ultimately, their commitment to ethical principles will enhance organizational resilience and contribute to the responsible development of AI technologies that benefit society as a whole. Through these efforts, ethical leaders not only advance their organizations' objectives but also contribute to a more just and equitable society in the face of technological transformation.



\begin{thebibliography}{}
\bibitem[Bastit et al., 2023]{R1}
Bastit B, Fernández M, Kesh S, Tsui D (2023) The AI Governance Challenge, 29 Nov.
{\url{https://www.spglobal.com/en/research-insights/special-reports/the-ai-governance-challenge}}
\bibitem[Brevini, 2020]{R2}
Brevini, B. (2020). Black boxes, not green: Mythologizing artificial intelligence and omitting the environment. Big Data \& Society, 7(2). {\url{https://doi.org/10.1177/2053951720935141}}
\bibitem[Brown and Treviño, 2006]{R3}
Brown ME and Treviño LK (2006) Ethical leadership: A review and future directions. The leadership quarterly, 17(6), 595-616. 
{\url{https://www.sciencedirect.com/science/article/abs/pii/S104898430600110X}}
\bibitem[Buck, 2023]{R4}
Buck B (2023) Building a Culture of Ethical Leadership | Infinite Strengths, 28 Dec.
{\url{ https://infinitestrengths.com/blog-ethical-leadership/}}
\bibitem[Ciulla and Ciulla, 2020]{R5}
Ciulla JB and Ciulla JB (2020). The importance of leadership in shaping business values. The search for ethics in leadership, business, and beyond, 153-163. {\url{https://link.springer.com/chapter/10.1007/978-3-030-38463-0_10
}}
\bibitem[Crews, 2015]{R6}
Crews J (2015) What is an ethical leader?: The characteristics of ethical leadership from the perceptions held by Australian senior executives Journal of Business and Management, 21(1), 29-58. 
\bibitem[Desimone, 2024]{R7}
Desimone D (2024) Why It’s Critical to Build Trust in an AI World Salesforce,31 July. {\url{https://www.salesforce.com/news/stories/helping-businesses-trust-ai/}}
\bibitem[Dignum, 2019]{R8}
Dignum V (2019) Responsible Artificial Intelligence: Designing AI for Human Values Proceedings of the 2019 AAAI\/ACM Conference on AI, Ethics, and Society. {\url{https://www.itu.int/en/journal/001/Documents/itu2017-1.pdf
}}
\bibitem[Eitel-Porter R, 2021]{R9}
Eitel-Porter R (2021) Beyond the promise: implementing ethical AI. AI Ethics 1, 73–80. {\url{https://doi.org/10.1007/s43681-020-00011-6
}}
\bibitem[Fountaine et al., 2019]{R10}
Fountaine T, McCarthy B, Saleh T (2019) Building the AI-Powered Organization Harvard Business Review. {\url{https://wuyuansheng.com/doc/Databricks-AI-Powered-Org__Article-Licensing-July21-1.pdf
}}
\bibitem[Frey and Osborne, 2017]{R11}
Frey CB and Osborne M (2017) The future of employment: how susceptible are jobs to computerisation? Technological Forecasting and Social Change, 114:254-280. {\url{https://www.sciencedirect.com/science/article/abs/pii/S0040162516302244
}}
\bibitem[Hartog, 2015]{R12}
Hartog D (2015) Ethical leadership Annu. Rev. Organ. Psychol. Organ. Behav., 2(1), 409-434. {\url{https://www.annualreviews.org/content/journals/10.1146/annurev-orgpsych-032414-111237
}}
\bibitem[Holdsworth, 2023]{R13}
Holdsworth J (2023) What is AI bias, 22 Dec. 
{\url{ https://www.ibm.com/topics/ai-bias}}
\bibitem[Judge et al., 2024]{R14}
Judge B, Nitzberg M, Russell S, When code isn’t law: rethinking regulation for artificial intelligence, Policy and Society, 2024;, puae020.
{\url{ https://doi.org/10.1093/polsoc/puae020}}
\bibitem[Kang, 2019]{R15}
Kang S-W (2019) Sustainable Influence of Ethical Leadership on Work Performance: Empirical Study of Multinational Enterprise in South Korea. Sustainability.11(11):3101.
{\url{ https://doi.org/10.3390/su11113101}}
\bibitem[Kaplan, 2016]{R16}
Kaplan J (2016) Artificial intelligence: What everyone needs to knowR. Oxford University Press. Mantelero A (2022) Beyond data: Human rights, ethical and social impact assessment in AI (p. 200) Springer Nature.
{\url{ https://library.oapen.org/handle/20.500.12657/57009}}
\bibitem[Mendonca and Kanungo, 2007]{R17}
Mendonca M and Kanungo R (2007) Ethical Leadership. United Kingdom: McGraw-Hill Education. 
\bibitem[Seldon, 2021]{R18}
Seldon (2021) Six Types of AI Bias Everyone Should Know, 12 Oct. 
{\url{ https://www.seldon.io/6-types-of-ai-bias}}
\bibitem[Tennin et al., 2024]{R19}
Tennin KL, Ray S, Sorg JM (Eds.) (2024). Cases on AI Ethics in Business. IGI Global. 
{\url{ https://doi.org/10.4018/979-8-3693-2643-5}}
\bibitem[UN environment programme, 2024]{R20}
UN environment programme (2024). AI has an environmental problem. Here’s what the world can do about that, 21 Sep.
{\url{ https://www.unep.org/news-and-stories/story/ai-has-environmental-problem-heres-what-world-can-do-about}}
\bibitem[Van Den Akker et al., 2009]{R21}
Van Den Akker L, Heres L, Lasthuizen KM, Six FE (2009). Ethical leadership and trust: It's all about meeting expectations. International Journal of leadership studies, 5(2), 102-122. 
{\url{ https://research.vu.nl/ws/portalfiles/portal/2548684/230514.pdf}}
\bibitem[Walmsley, 2021]{R22}
Walmsley, J. Artificial intelligence and the value of transparency. AI \& Soc 36, 585–595 (2021). 
{\url{https://doi.org/10.1007/s00146-020-01066-z}}
\bibitem[Yaou and Hyounae, 2023]{R23}
Yaou Hu and Hyounae (Kelly) Min (2023) The dark side of artificial intelligence in service: The “watching-eye” effect and privacy concerns", 10 Feb. 
{\url{https://www.sciencedirect.com/science/article/abs/pii/S0278431923000117}}

\end{thebibliography}
\end{document}